# Escape of an active ring from an attractive surface: behaving like a self-propelled Brownian particle

Bin Tang, Jin-cheng Gao, Kang Chen[§], Tian Hui Zhang[†], Wen-de Tian*

Center for Soft Condensed Matter Physics and Interdisciplinary Research, Soochow University, Suzhou 215006, China

* tianwende@suda.edu.cn;  [†]zhangtianhui@suda.edu.cn;  [§]kangchen@suda.edu.cn

**Abstract**

Escape of active agents from metastable states is of great interest in statistical and biological physics. In this study, we investigate the escape of a flexible active ring, composed of active Brownian particles, from a flat attractive surface using Brownian dynamics simulations. To systematically explore the effects of activity, persistence time, and the shape of attractive potentials, we calculate escape time $\tau_e$ and effective temperature $T_{\text{eff}}$. We observe two distinct escape mechanisms: Kramers-like thermal activation at small persistence times, where $\ln(\tau_e) \sim 1/(k_B T_{\text{eff}})$, and the maximal force problem at large persistence time, where $\tau_e$ is determined by persistence time. The escape time explicitly depends on the shape of the potential barrier at high activity and large persistence time. Moreover, when the propulsion force is biased along the ring's contour, escape becomes more difficult and is primarily driven by thermal noise. Our findings highlight that, despite its intricate configuration, the active ring can be effectively modeled as a self-propelled Brownian particle when studying its escape from a smooth surface.

DOI:

## 1. Introduction

The escape of a Brownian particle over a potential barrier is a thermally activated process, a classic problem known as Kramers' problem [1,2]. Kramers' theory [3,4] suggests that in the limit of vanishing particle flux across a potential barrier, the escape rate decreases exponentially with the increase in barrier height. Recently, active particles, which can convert environmental energy into directed motion, have attracted significant attention because they are inherently out of equilibrium [5–8]. Due to self-propulsion, active particles are expected to escape a potential barrier at a higher rate than passive particles. Activated escapes play a vital role in many physical phenomena, such as the nucleation in motility-induced phase separation [9] and escape through narrow channels [10]. Additionally, in dense phases, where each particle is confined by its neighbors, the dynamics are controlled by the rate at which particles undergo reorganization and overcome local energy barriers [11]. Although the coupling of orientational and positional degrees of freedom in active particles makes the theoretical treatment of the escape problem challenging [12–14], theorists have made significant efforts in both near-equilibrium and far-from-equilibrium regimes [15–18]. For instance, Woillez et al. [19] found that the escape time of self-propelled Brownian particles is affected not only by the overall shape of the potential, but also by the height of the potential barrier. Caprini et al. [20,21] discovered that the velocity distribution of active particles around the potential barrier exhibits a bimodal shape in large persistence regimes.

In addition to point-like active particles, there exist chain- or filament-like structures, usually termed active polymers [22], such as filamentous actins or microtubules of the cell cytoskeleton propelled by tread-milling and motor proteins [23]. Computer simulation [24,25] and theoretical studies [26–28] have shown that flexible linear active Brownian polymers swell with increasing activity, and their overall diffusive dynamics are enhanced. Active Brownian rings have also received considerable attention due to their relevance in biological systems [29]. For example, the properties of ring-like DNA molecules are affected by ATP-dependent enzymatic activity, which induces mechanical fluctuations in the cytoplasm of eukaryotic cells [30]. Dynamically, the mean square displacement of an active Brownian ring exhibits ballistic, sub-diffusive, and diffusive regimes, indicating a distinct activity-enhanced diffusion [29].

An interesting question arises: how does the escape of an active Brownian ring behave? The motivation is threefold: First, from a physical perspective, tightly linked active particles are associated with persistent active stresses, unlike individual active particles [27]. These stresses alter the conformational and dynamical properties of active polymers, thereby affecting their escape behaviors. Second, the escape behavior of biopolymers in confined environments is crucial for the survival and development of organisms [31]. Third, a typical escape phenomenon in polymer physics is the interfacial adsorption and desorption, which has numerous applications in industry and biomedicine [32]. As is well known, the complete escape of a polymer is unlikely because it is improbable that all segments will detach simultaneously. To promote escape, the method of "increasing temperature" is often employed [33], where escape is driven by the increasing configurational entropy of the chain, counteracting the enthalpy from polymer-surface interactions. In contrast to thermal noise, the "active" force introduces activity to polymers. However, how this "activity" influences escape behavior remains an open question.

In this study, we investigate the escape of an active Brownian ring (ABR) from an attractive surface. We define an effective temperature based on the position fluctuations of the ABR's center of mass. Our results reveal that the escape time exhibits an exponential dependence on the inverse of the effective temperature at moderate activities, similar to Kramers' thermal activation mechanism. At higher activities, the escape time is inversely proportional to active force because it is shorter than the crossover time between ballistic and diffusive motion. Additionally, we examine the effects of the persistence time $\tau_R$ of active Brownian particles and the potential barrier. The critical barrier height depends non-monotonically on persistence time due to the presence of two distinct escape mechanisms at small and large $\tau_R$: thermal activity and the maximal force problem. Furthermore, we find that at small $\tau_R$, the escape time is weakly dependent on the barrier shape. However, at large $\tau_R$, the shape plays a crucial role in escape dynamics, since it determines the maximal force that the ring must overcome. Moreover, we observe that the driven mode also influences the escape. When the active force is biased along the ring's contour, escape becomes more difficult and is primarily driven by background noise. Our findings suggest that the escape of ring-like structures can be understood similarly to that of an active Brownian particle.

## 2. Model and methods

We perform Brownian dynamics simulations to investigate the behavior of a ring-like chain near an attractive, homogeneous surface. The ring is composed of $N$ active Brownian monomers connected by a bead-spring model commonly used in polymer physics. For an active Brownian ring (ABR), the intrinsic orientation $\hat{q}$ of each monomer is initially random. A propulsive force $\boldsymbol{F_a} = F_a \hat{q}$ is applied along $\hat{q}$, as shown in Fig. 1(a), where $F_a$ has a constant magnitude, following the form typically used for active Brownian particles [7].

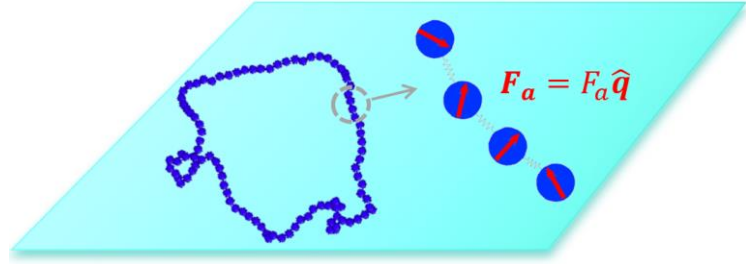

FIG. 1: Schematic illustration of an active Brownian ring (ABR) adsorbed on an attractive surface. The arrows indicate the directions of the active forces exerted on each monomer, which follow the same form as those in active Brownian particles.

In our model, the total interaction $U_{tot}$ is divided into three parts: the interaction between monomers $U_{WCA}$, the bond interaction between adjacent monomers $U_b$, and the attractive interaction $U_s$ by surface.

$$U_{tot} = U_{WCA} + U_b + U_s \quad (1)$$

Here $U_{WCA} = \sum_{i<j}^{N} U(r_{ij})$ is represented by the pure repulsive WCA potential given below

$$U(r_{ij}) = \begin{cases} 4\varepsilon\left[\left(\frac{\sigma}{r_{ij}}\right)^{12} - \left(\frac{\sigma}{r_{ij}}\right)^{6}\right] + \varepsilon & r_{ij} \leq 2^{1/6}\sigma \\ 0 & \text{otherwise} \end{cases} \quad (2)$$

where $r_{ij}$ represents the distance between the $i$th and $j$th monomers. $\varepsilon$ denotes the interaction strength and $\sigma$ is the diameter of the monomer. The bond interaction $U_b$ is expressed as $U_b = \frac{1}{2}k\sum_{i=1}^{N-1}(b_i - b_0)^2$ with spring coefficient $k = 2000.0 k_B T/\sigma^2$. $b_i$ denotes the $i$th bond length and $b_0 = \sigma = 1.0$ is the equilibrium bond length.

To investigate the impact of shape of potential barrier $U_s = \sum_{i=1}^{N} U(z_i)$ on escape, three potential functions ($U_1$, LJ-9-3; $U_2$, harmonic; $U_3$, Morse) were adopted with parameters adjusted to ensure the same height of potential barriers at a truncation distance $z_c = 3.0\sigma$. The details of these functions are listed below.

$$U_1(z_i) = \begin{cases} \varepsilon_{s1}\left[\frac{2}{15}\left(\frac{\sigma}{z_i}\right)^9 - \left(\frac{\sigma}{z_i}\right)^3\right] & z_i \leq z_c \\ 0 & \text{otherwise} \end{cases}$$

$$U_2(z_i) = \begin{cases} \varepsilon_{s1}\left[\frac{2}{15}\left(\frac{\sigma}{z_i}\right)^9 - \left(\frac{\sigma}{z_i}\right)^3\right] & z_i \leq 0.86\sigma \\ -\varepsilon_{s2}(z_i - z_c)^2 & 0.86\sigma \leq z_i \leq z_c \\ 0 & \text{otherwise} \end{cases} \quad (3)$$

$$U_3(z_i) = \begin{cases} \varepsilon_{s1}\left[\frac{2}{15}\left(\frac{\sigma}{z_i}\right)^9 - \left(\frac{\sigma}{z_i}\right)^3\right] & z_i \leq 0.86\sigma \\ \varepsilon_{s3}\left[e^{-3.72(z_i - z_0)/\sigma} - 2e^{-1.86(z_i - z_0)/\sigma}\right] & 0.86\sigma \leq z_i \leq z_c \\ 0 & \text{otherwise} \end{cases}$$

where $\varepsilon_s s$ (including $\varepsilon_{s1}$, $\varepsilon_{s2}$, and $\varepsilon_{s3}$) are the interaction strength between a monomer and the surface. $z_i$ is the distance between the $i$th monomer and the surface. $z_0 = 0.86\sigma$ is the lowest point of the Morse potential. It should be pointed out that the three potentials have the same repulsive parts for $z_i \leq 0.86\sigma$; the difference lies in their attractive parts.

The motion of each monomer is described by the overdamped Langevin equations,

$$\gamma_T \frac{dr(t)}{dt} = -\nabla U_{tot} + F_a \hat{q}(t) + \sqrt{6\gamma_T^2 D_T} \boldsymbol{\eta}^T \quad (4)$$

$$\frac{d\hat{q}(t)}{dt} = \sqrt{4D_R} \cdot \boldsymbol{\eta}^R \times \hat{q}(t) \quad (5)$$

where $\gamma_T$ denotes translational friction coefficient, $D_T(=\frac{k_BT}{\gamma_T})$ and $D_R$ are translational and rotational diffusion coefficients of the monomers, respectively. $\boldsymbol{\eta}^T$ and $\boldsymbol{\eta}^R$ are independent Gaussian noises with zero mean and unit variance, with the correlations $\langle \eta_\mu^T(t)\eta_\nu^T(t')\rangle = \langle \eta_\mu^R(t)\eta_\nu^R(t')\rangle = \delta_{\mu\nu}\delta(t-t')$, here $\mu, \nu$ are the components of translational or rotational degrees of freedom. Generally, for spherical monomers, $D_T/D_R = \sigma^2/3$, but not obeyed in our simulation. It is possible that $D_R$ is completely independent of $D_T$ [34]. Hence, we keep $D_T = 1.0$ and vary $D_R$ as a parameter to study its influence. The persistence time, $\tau_R = 1/(2D_R)$, represents the time needed for the active force of each monomer to change its direction.

We used a home-modified LAMMPS software to perform simulations of all systems. The equations of motion were solved by velocity-Verlet algorithm. An ABR was placed in a box of $100\sigma \times 100\sigma \times 50\sigma$. Periodic boundary conditions were applied in $x$- and $y$- directions, the fixed boundary condition in $z$-direction. A repulsive wall at the top of box was used to hold back the ring. $N$=100, $T$=1.0, $\gamma_T = 1.0$, and a time step $dt = 10^{-4}\tau$ with $\tau = \sigma^2/D_T$. The dimensionless Péclet number (Pe) was defined as $Pe = F_a\sigma/k_BT$ to characterize the activity. Each simulation was run for a long duration ($10^4\tau$) to ensure adequate data collection.

To analyze the structure and dynamics of the ring, we define several quantities: $Rg_\parallel^2(t)$ and $Rg_\perp^2(t)$, which are the components of the mean square radius of gyration; $MSD_\parallel(t)$ and $MSD_\perp(t)$, which are the components of the mean square displacement (MSD) of the center of mass; the detached number of monomers $M(t)$; and the correlation function $G_c(t)$. Details are given below and the Supplemental Material [35].

$$x_{cm}(t) = \frac{1}{N}\sum_{i=1}^{N} x_i(t), y_{cm}(t) = \frac{1}{N}\sum_{i=1}^{N} y_i(t), z_{cm}(t) = \frac{1}{N}\sum_{i=1}^{N} z_i(t)$$

$$Rg_{xy}^2(t) = \frac{1}{N}\sum_{i=1}^{N}\{[x_i - x_{cm}]^2 + [y_i - y_{cm}]^2\}, Rg_z^2(t) = \frac{1}{N}\sum_{i=1}^{N}\{[z_i - z_{cm}]^2\}$$

$$MSD_{xy}(t) = \langle (x_{cm}(t) - x_{cm}(0))^2 + (y_{cm}(t) - y_{cm}(0))^2 \rangle \quad (6)$$

$$MSD_z(t) = \langle (z_{cm}(t) - z_{cm}(0))^2 \rangle$$

$$G_c(t) = \langle \frac{N - M(t)}{N - M(0)} \rangle$$

where $r_i(t) = [x_i(t), y_i(t), z_i(t)]$ and $r_{cm}(t) = [x_{cm}(t), y_{cm}(t), z_{cm}(t)]$ represent the coordinates of the $i$th monomer and the center-of-mass of the ring at time $t$, respectively. The notation $\langle \cdots \rangle$ denotes the ensemble average over 1000 independent simulations. The average number of escaped monomers, $\langle M \rangle$, is also calculated [35]. The escape time, $\tau_e$, is determined by fitting the correlation function $G_c(t) \sim exp(-t/\tau_e)$. Note that re-adsorbed monomers are not counted when calculating $G_c(t)$. We systematically analyze the structure and dynamics of rings with the potential barrier $U_1(z_i)$ at $\varepsilon_{s1} = 5.0$ and present the results below, unless otherwise specified.

## 3. Results

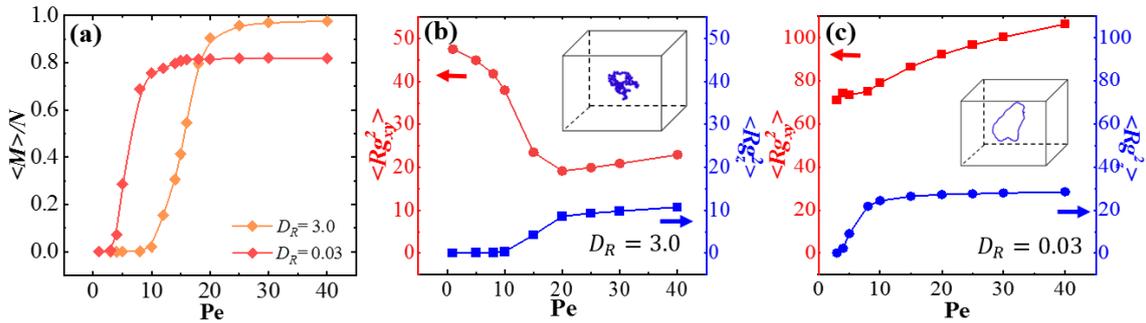

FIG. 2. (a) The average numbers of escaped monomers $\langle M \rangle$ vary with Pe for $D_R = 3.0$ and 0.03. (b) The $xy$ (red)- and $z$ (blue)- components of radius of gyration as a function of Pe for $D_R = 3.0$. (c) The $xy$ (red) - and $z$ (blue)- components of radius of gyration as a function of Pe for $D_R = 0.03$. Insets are snapshots for Pe=25, with blue and red arrows provided for visual guidance.

*Effect of activity.* We first focus on the structure of ABR near the surface as the activity, Pe, is varied through changes in the propulsive force $F_a$. The $\langle M \rangle$s as a function of Pe for $D_R$=3.0 and 0.03 are shown in Fig.2a. For $D_R = 3.0$, at low activity, all monomers are adsorbed on the surface. At high activities, almost no monomers remain adsorbed, indicating the escape of the ABR (see FIG.S1 in the Supplemental Material [35] for $M(t)$). The behavior is qualitatively similar for $D_R = 0.03$, except that Pe required for ABR escape becomes smaller. Thus, we conclude that increased activity promotes the escape of ABRs from the attractive surface. This effect is further illustrated by the components of $\langle Rg^2 \rangle$ as a function of Pe, shown in Fig. 1b and Fig. 1c (also in Fig.S2) for $D_R=3.0$ and 0.03, respectively. For $D_R=3.0$, $\langle Rg_{xy}^2 \rangle$ initially decreases and then slightly increases with increasing Pe, in good agreement with the change of chain size in three dimensions with volume-excluded interactions[17]. Additionally, $\langle Rg_z^2 \rangle$ shows a monotonic increase with increasing Pe. In contrast, for $D_R=0.03$, $\langle Rg_{xy}^2 \rangle$ monotonically increases with Pe, accompanying the ABR escape. This suggests that the escape of the ring is associated with conformational fluctuations that depend on both the activity and the rotational diffusion coefficient.

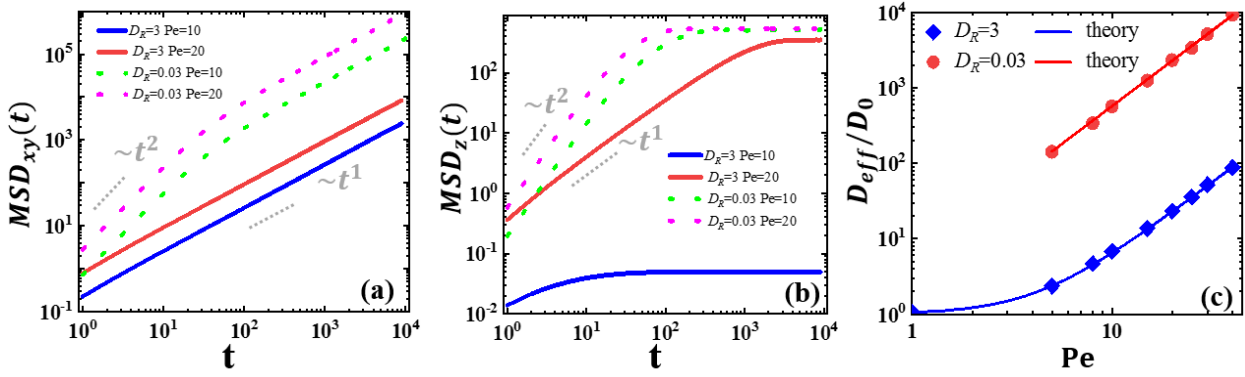

FIG.3.Time evolution of (a) $MSD_{xy}(t)$ and (b) $MSD_z(t)$. Dotted lines indicate the normal and super-diffusive exponents. (c) The effective diffusion coefficients as a function of Pe. The curves represent theoretical results: $D_{\text{eff}} = D_0(1 + aPe^2)$, with $D_0 = \frac{k_BT}{\gamma_T N}$ and $a = \frac{D_T^2}{\sigma^2 D_R}$.

Now we turn to the center-of-mass diffusion of ABR via calculating the $MSD_{xy}(t)$ and $MSD_z(t)$, given in Fig.3a and Fig.3b. For $D_R = 3.0$, normal diffusion is observed in the *xy*-direction. When Pe < 10, sub-diffusion occurs in the *z*-direction due to the ABR being adsorbed on the surface. When Pe ≥ 20, the ABR exhibits normal diffusion in the *z*-direction as it escapes from the surface. For $D_R = 0.03$, super-diffusion is observed at short time scales, while normal diffusion is observed at long time scales in the *xy*-directions for all *Pe*s, which is consistent with theoretical analysis [36]. The center-of-mass $MSD(t)$ in the bulk can be described by $MSD(t) = \frac{2dk_BT}{N\gamma_T}t + \frac{2v_0^2}{N(2D_R)^2}[(2D_R)t - 1 + e^{-(2D_R)t}]$ [36], where $d = 3$ is the dimension. Consequently, $MSD_s(t) = \frac{2dk_BT}{N\gamma_T}t + \frac{v_0^2}{N}t^2$ at short time scales and $MSD_l(t) = (\frac{2dk_BT}{N\gamma_T} + \frac{v_0^2}{ND_R})t$ at long time scales, where $v_0 = P_e D_T/\sigma$ is the propulsion speed. The crossover time between ballistic and diffusive motion is $\tau_c = 1/D_R$ (derived from $MSD_s(t) = MSD_l(t)$). Using Einstein's relation $MSD_{xy}(t) = 4D_{\text{eff}}t$, the effective diffusion coefficient of the ring is obtained (Fig.3c). The effective diffusion coefficient yields the function $D_{\text{eff}} = D_0(1 + aPe^2)$ with $a = \frac{D_T^2}{\sigma^2 D_R}$ and $D_0 = \frac{k_BT}{\gamma_T N}$. For $D_R = 3.0 = 3D_T$, $a$ is close to the 1/18, as predicted theoretically where steric interactions are omitted [36]. Fig. 3c shows that the simulated $D_{\text{eff}}$s agree well with theoretical predictions. According to the position fluctuations of ring's center-of-mass, one could define an effective temperature, $T_{\text{eff}}$, replacing the thermal temperature in describing ABR' properties. The effective temperature is given by $T_{\text{eff}} = T(1 + aPe^2)$, where $T$ is the thermal noise. The significance of $T_{\text{eff}}$ is twofold: First, each monomer of the ABR experiences an active (non-thermal) force, making the concept of temperature from equilibrium insufficient to describe the ABR. Second, it establishes the relation between $T_{\text{eff}}$ and activity, which helps in understanding non-equilibrium systems by utilizing concepts from equilibrium states.

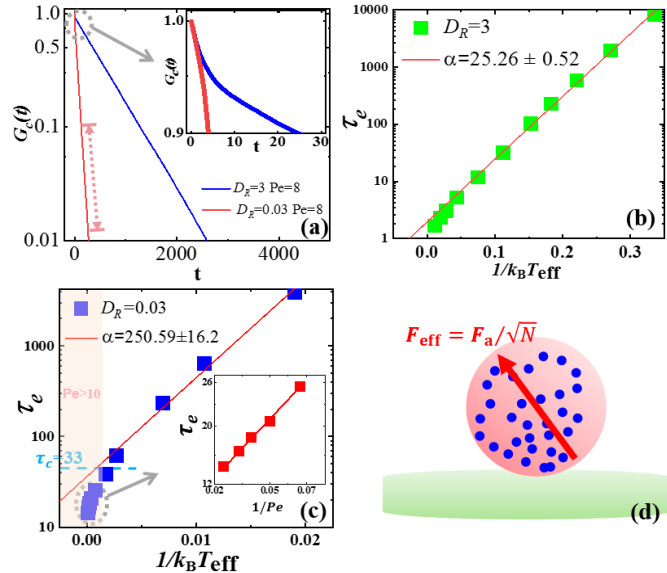

FIG. 4. (a) Decay of $G_c(t)$ over time. The dashed line is a guide to the eye for the fitting regime where $G_c(t) \sim exp(-t/\tau_e)$. The inset shows the rapid escape of monomers at a short transient period. (b) and (c) Semi-logarithmic plot of escape time $\tau_e$ as a function of $1/k_B T_{eff}$ for $D_R = 3.0$ and 0.03, respectively. The red lines represent fits to the Kramers-like expression $\tau_e \sim exp(\alpha/k_B T_{eff})$. The inset in (c) shows $\tau_e$ as a function of $1/Pe$ for $D_R = 0.03$ at large activities. The crossover time $\tau_c = 1/D_R = 33$ for $D_R = 0.03$ is also indicated. (d) Sketch of an active Brownian particle illustrating the escape mechanism of an active Brownian ring from a surface. The effective propulsion force experienced by the particle is $F_a/\sqrt{N}$.

Semi-logarithmic plot of $G_c(t)$ for various Pes is shown in Fig.4a. There is a short transient period where monomer escape occurs rapidly, followed by an exponential decay at longer timescales. The escape time $\tau_e$ is obtained by fitting $\ln(G_c(t)) \sim -t/\tau_e$. To understand the escape mechanism induced by the activity, the dependence of $\tau_e$ on $1/k_B T_{eff}$ is plotted semi-logarithmically in Fig. 4b and Fig.4c for $D_R = 3.0$ and 0.03, respectively. For large $D_R (= 3.0)$, the escape time follows a Kramers-like expression, $\tau_e \sim exp(\alpha/k_B T_{eff})$, with a fitting parameter $\alpha \approx 25$. Here $\alpha$ is relevant to the average adsorption energy of the ring. This indicates that the escape mechanism is similar to that of passive Brownian particles at an effective temperature $T_{eff}$. For small $D_R (= 0.03)$, there are evidently two regimes: for $Pe < 10$, $\tau_e$ fits well with $\tau_e \sim exp(\alpha/k_B T_{eff})$ with $\alpha \approx 230$; for $Pe > 10$, $\tau_e$ linearly increases with $1/Pe$. This indicates that, in high activity regimes, the escape mechanism transitions from thermal activation to a ballistic-like escape. The ring behaves similarly to an active Brownian sphere. The effective thermal diffusion coefficient of the sphere is $D_T/N$, and its effective propulsion velocity is $v_0/\sqrt{N}$. The rotational diffusion coefficient of the sphere is the same to that of a monomer. For $\tau_e > \tau_c$, the escape process resembles a thermally activated process, while for $\tau_e < \tau_c$, it exhibits ballistic behavior. This physical picture aligns with the findings presented here.

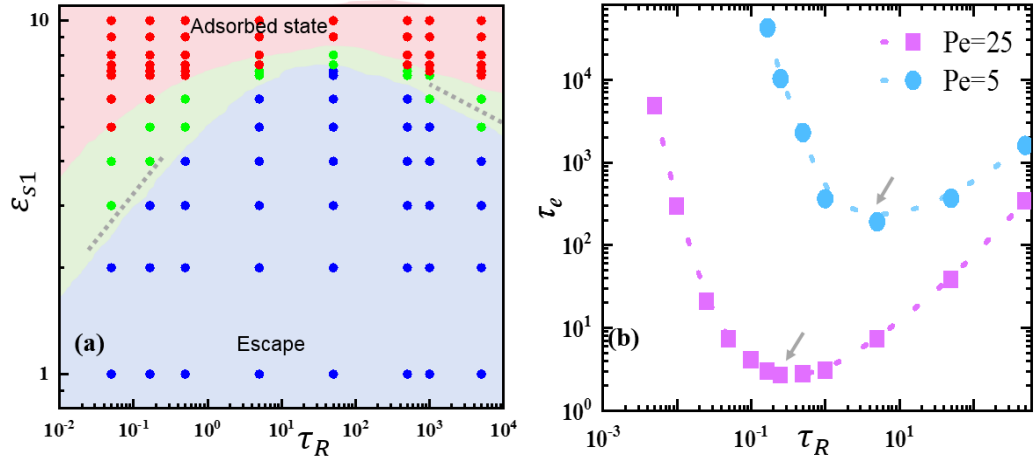

FIG. 5 (a) Phase diagram in the $\varepsilon_{s1} - \tau_R$ plane for Pe=5. $\tau_R = 1/(2D_R)$ is the persistence time of a monomer. Red points denote adsorbed states, the blue escape, and the green intermediate regime. The dashed lines are for eye-guiding phase boundary (b) $\tau_e$ as a function of $\tau_R$ for Pe = 25 and 5. The arrows point to the optimal persistence time.

**Phase diagram.** $\tau_e$s are different at various $D_R$s, even with the same potential barrier and activity. To better understand the influence of $D_R$, a phase diagram is shown in Fig.5a for Pe=5. Here, we use $\tau_R$ instead of $D_R$ to present the phase diagram more clearly. Two states are distinguished by $\tau_e$. For an escape state, $\tau_e$ is less than our simulation time ($10^4 \tau$). For an adsorbed state, $\tau_e$ is too larger to be fitted. Additionally, an intermediate regime is defined when $\tau_e$ can be fitted but is larger than our simulation time. At a certain activity, the high barrier leads to adsorbed states, while for low barriers, escape occurs regardless of $\tau_R$ due to the existence of background noise. It is noted that, due to the counteracting effects of activity and attractive potential, larger $\varepsilon_{s1}$s are required for the appearance of adsorbed states at higher Pes. However, the dependence of the phase boundary shape on $\tau_R$ might not change.

At the phase boundary, the critical barrier first increases and then decreases with the increase of $\tau_R$. This behavior arises from different escape mechanisms at small and large $\tau_R$s, also manifested by $\tau_e$ as a function of $\tau_R$. As shown in Fig.5b, $\tau_e$ decreases initially and then increases with $\tau_R$. For small $\tau_R$s, the monomers rotate quickly, and the whole ring behaves like a passive system with an effective temperature $T_{eff}$, so $\ln \tau_e \sim \alpha/\tau_R$. This explains why $\tau_e$ decreases as $\tau_R$ grows at first. For large $\tau_R$s, escape occurs almost deterministically when the magnitude of the active force exceeds the maximal force imposed by the potential barrier. Since the active force has little chance of reversing orientation while crossing the barrier, the escape time approximates the waiting time for the occurrence of the driven force in the direction perpendicular to the surface that exceeds the maximal force. The barrier crossing becomes the maximal force problem, where $\tau_e$ is determined by $\tau_R$. As a result, a larger $\tau_R$ implies a longer escape time [37]. Additionally, the remarkable non-monotonic behavior demonstrates that, for a given Pe, there exists an optimal $\tau_R$. The optimal $\tau_R$ can be roughly estimated by the ratio of barrier width and propulsive velocity, meaning that it is inversely proportional to activity. Indeed, the optimal $\tau_R$ at Pe=10 is smaller than that at Pe=5 (Fig.5b). The finding qualitatively agrees with that of Caprini et al. [21], who studied the escape of an active Brownian particle from a double-well potential.

The non-monotonicity of critical barriers at the same activity can be explained below. In the regime of small $\tau_R$s, activated noise dominates. Critical barriers are estimated by $\approx k_B T_{\text{eff}}$. Since $T_{\text{eff}} \propto \frac{v_0^2}{ND_R} \propto \tau_R$, the critical barriers increase with $\tau_R$. In the regime of large $\tau_R$s, the maximal force dominates. A larger $\tau_R$ means a longer waiting time for the driven force perpendicular to the surface to reach its maximum. Therefore, with the limited waiting time (our simulation time is $10^4\tau$) and the fixed Pe, critical barrier decreases as $\tau_R$ increases. It should be pointed out that the non-monotonic transport behavior of active particles [38] and active polymers [39] under confinement has also been observed recently. This behavior arises from the competition between two length scales: the persistence length (determined by activity and reorientation dynamics) and the geometrical length due to confinement boundaries or obstacles [38,39]. The mechanism is general for the navigation of active agents through disordered, porous environments and is akin to ours. The difference here is that the geometrical length is determined by the potential barrier.

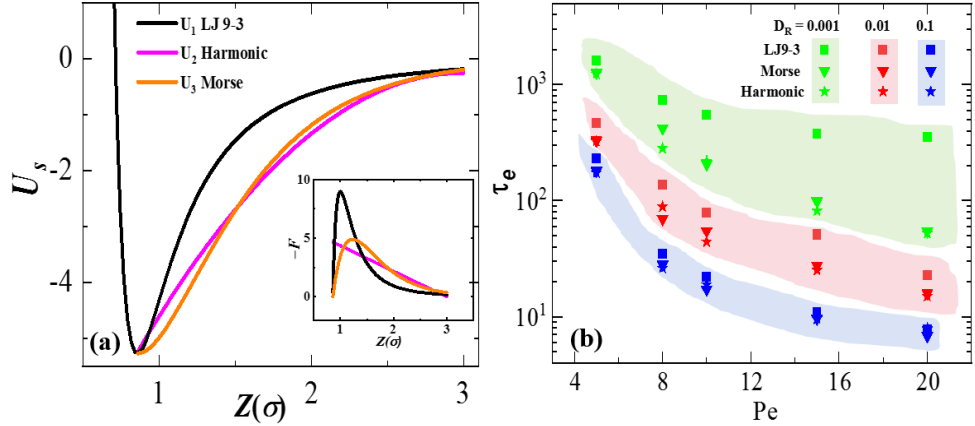

FIG. 6. (a) The shapes of the three barriers: Harmonic potential, LJ9-3 potential, and Morse potential with $\varepsilon_{s1} = 5.0, \varepsilon_{s2} = 1.09,$ and $\varepsilon_{s2} = 5.27$. The inset is the corresponding force. (b) The escape times as functions of Pe for the three barriers with various $D_R$s.

***Effect of barrier shape.*** Now we are interested in the effect of the shapes of the potential barriers on ABR's escape. To realize the different potential shapes, the attractive region of LJ-9-3 potential ($U_1$) was replaced by that of harmonic ($U_2$) and Morse potential ($U_3$). The shapes of potentials and forces are plotted in Fig.6a. To study the effect of barrier shapes, we kept the heights of three barriers nearly the same, $\Delta U \approx 5.0 k_B T$, in the range of z from $0.86\sigma$ to $3.0\sigma$. The escape time as a function of activity for three $D_R$s (=0.1, 0.01, 0.001) is presented in Fig.6b. For large $D_R$(=0.1, blue background), the escape times for the three barriers are very close for all Pe values, implying that the escape weakly depends on the barrier shapes because the thermally activated mechanism is dominant here. The behavior is similar for small Pe (=4.0). At small $D_R$(=0.001, blue background) and large Pes, the deviation in escape time becomes significantly large, especially for the LJ-9-3 potential. This indicates that the barrier shape indeed impacts the escape time of ABR when the persistence length of monomers is large, similar to the findings of Woillez et al. [19], who studied the noise-driven escape of active Brownian particles. The reason might be that the maximal trapping force exerted by different barrier shapes on the monomer varies. The maximal trapping force of LJ-93 potential is larger than that of the other two potentials (see the inset of Fig.6a).

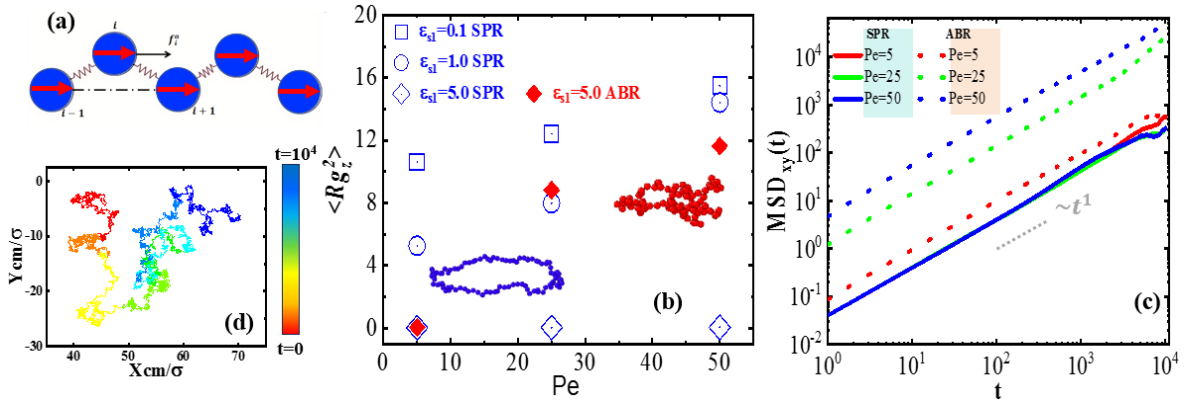

FIG. 7. (a) Sketch of a self-propelling ring (SPR). (b) $<Rg_z^2>$ as a function of Pe for various $\varepsilon_{s1}$s. The $<Rg_z^2>$ of ABR at $\varepsilon_{s1} = 5.0$ is also given for comparison. The insets show typical snapshots of SPR (blue beads) and ABR (red beads) at $\varepsilon_{s1} = 5.0, \text{Pe} = 50$ (c) MSD of SPR and ABR for various activities at $\varepsilon_{s1} = 5.0$. (d) The time evolution of trajectory for the center-of-mass of SPR.

***Effect of propulsive mode.*** For the active Brownian ring, the orientations of monomers do not correlate with the chain conformation, so the entire chain behaves like an active Brownian sphere. An interesting question is how the coupling of propelling directions with chain configuration affects escape behavior. To explore this, we now restrict the orientations of the monomers to align with the contour of the ring, i.e., biasing the orientations to preferentially align with the instantaneous tangent vector of the ring's contour through a potential $\frac{K}{k_BT}(\hat{n}_i - \hat{b}_i)^2$, were $\hat{b}_i = \frac{r_{i+1,i-1}}{|r_{i+1,i-1}|}$ is the $i$th vector, and $K$ is the biasing strength. In the following, we focus on the case of strong restriction ($K \to \infty$), that is often called the self-propelling ring (SPR), a sketch of which is given in Fig.7a.

The $\langle Rg_z^2 \rangle$s as a function of Pe are shown in Fig.7b. The value of $\langle Rg_z^2 \rangle$ close to zero means the SPR is adsorbed on the surface, otherwise, it has escaped from the surface. For $\varepsilon_{s1} = 5.0$, the $\langle Rg_z^2 \rangle$ of the ABR is also provided for comparison. It can be found that, the escape of the SPR occurs at the attractive strengths near or below $\varepsilon_{s1} \leq 1.0$ (see Fig.S3 [35] for the distribution of monomers along the $z$ direction). The ABR escapes relatively easily at the same activity levels and attractive barriers, implying that the propulsive mode impacts the escape dynamics of the ring. To understand why the SPR is difficult to escape from the flat surface, we calculate the $MSD_{xy}(t)$ of its center of mass at various activities. As seen in Fig.7c, increasing the magnitude of the active force does not seem to raise the effective diffusion coefficients. This behavior differs significantly from that of the ABR, whose effective diffusion coefficient increases with active force. This is also evident in the time evolution of the center-of-mass position of the SPR (Fig.7d), which shows diffusive motion on the surface. This implies that the escape of the SPR is primarily driven by environmental noise with a strength $T$. To further understand the finding, we theoretically derive the resultant active force on the center of mass of the SPR (see ref. [35] for details). The resultant force is proportional to $F_a R_E$, which is zero because the end-to-end vector $R_E = 0$ for a ring-like chain. Hence, the $MSD(t) = \frac{2dD_T}{N}$ irrelevant to the active force. This implies that the escape of the SPR is similar to that of passive particles, i.e., active Brownian particles with Pe = 0.

**4. Summary and discussion**

We explored the escape of a flexible active Brownian ring on a flat, attractive surface using computer simulations. At low activity, the ABR remains in an adsorbed state, diffusing along the surface. High activity, however, leads to the escape of the active ring. The diffusion of the ABR parallel to the surface aligns well with theoretical predictions. Consequently, an effective temperature, $T_{\text{eff}} = T(1 + aPe^2)$, can be defined due to the presence of non-thermal active noise. The escape kinetics were measured by the escape time $\tau_e$, determined by fitting $G_c(t) \sim exp(-t/\tau_e)$. Our results reveal that the escape time and the effective temperature follow the relationship $\tau_e \sim exp(\alpha/k_B T_{\text{eff}})$ at moderate activity, reminiscent of Kramers-like thermal activation. Interestingly, we also observed deviations from the exponential decay at higher activity due to the escape time being shorter than the crossover time, where the escape resembles a constant-speed "running away" mechanism.

Furthermore, we investigated the effects of persistence time $\tau_R$. We found a non-monotonic dependence of escape time on $\tau_R$ and identified an optimal $\tau_R$ (or $D_R$) for chain escape. We also focused on the shape of potential barriers and discovered that at large $D_R$ (small $\tau_R$), the escape time weakly depends on the shape, but at small $D_R$ (large $\tau_R$), the shape plays a crucial role in escape dynamics, as it determines the maximal force the ring must overcome. Additionally, we found that the mode of propulsion also affects escape behavior. For self-propelling rings, escape is difficult and primarily induced by background noise because the resultant force of activity on the center of mass is proportional to the end-to-end vector, which is zero for the ring.

As a conclusion, we propose that the active ring can be modeled as an active Brownian particle when studying its escape from a smooth surface. This mapping is helpful for understanding the effect of chain length $N$. A large $N$ only reduces the effective thermal diffusion coefficient and the effective propulsive velocity in the regime of Kramers-like thermal activation, resulting in an increase in escape time with $N$. However, if the active force easily overcomes the barrier—i.e., beyond the maximal force—the escape time becomes less dependent on chain length and is primarily determined by the persistence time $\tau_R$ [40].

It should also be noted that this study is a first step toward understanding activity-induced escape of chain-like active structures. Another intriguing question is the effect of inertia on each monomer, which could be explored using underdamped Langevin dynamics. Although our findings suggest that ring-like structures can be considered as active Brownian particles when studying their escape from a smooth surface, their escape from a cavity or passage through a pore with geometric confinement might differ significantly due to configuration entropy, making it a topic worth further study.

*Acknowledgments*. Project supported by the National Natural Science Foundation of China of Grant Nos. 21674078 (Tian), 21774091 (Chen), and 21574096(Chen).

----------------------------------------------------------------------